\documentclass[superscriptaddress,aps,pra,twocolumn,showpacs,nofootinbib,longbibliography]{revtex4-1}
\usepackage{algorithm}
\usepackage{algorithm}
\usepackage{balance}
\usepackage{algpseudocode,caption}
\usepackage{array}
\usepackage{amsmath,amsthm}
\usepackage{braket}
\usepackage{amsmath}
\usepackage{amsfonts}
\usepackage{amssymb}\usepackage[T1]{fontenc}
\usepackage{latexsym}
\usepackage{amssymb}
\usepackage[colorlinks=true,citecolor=blue,urlcolor=blue]{hyperref}
\usepackage{color}
\usepackage{graphics,epstopdf}
\usepackage{soul}
\usepackage[demo]{graphicx}
\usepackage{subcaption}
\usepackage{lipsum}
\usepackage{float}
\usepackage{adjustbox}
\usepackage[normalem]{ulem}

\newcommand{\be}{\begin{equation}}
\newcommand{\ee}{\end{equation}}
\newcommand{\ba}{\begin{eqnarray}}
\newcommand{\ea}{\end{eqnarray}}

\begin{document}
	
\title{Cryptanalysis of quantum permutation pad}

\author{Avval Amil}
\email{avval2801@gmail.com}
\affiliation{Department of Computer Science and Engineering, I.I.T. Delhi, Hauz Khas, New Delhi - 110016, India}

\author{Shashank Gupta}
\email{shashank@qnulabs.com}
\affiliation{QuNu Labs Pvt. Ltd., M. G. Road, Bengaluru, Karnataka 560025, India}

\date{\today}
\begin{abstract}

Cryptanalysis increases the level of confidence in cryptographic algorithms. We analyze the security of a symmetric cryptographic algorithm - quantum permutation pad (QPP) \cite{Kuang20}. We found the instances of ciphertext same as plaintext even after the action of QPP with the probability $\frac{1}{N}$ when the entire set of permutation matrices of dimension $N$ is used and with the probability $\frac{1}{N^m}$ when an incomplete set of $m$ permutation matrices of dimension $N$ are used. We visually show such instances in a cipher image created by QPP of 256 permutation matrices of different dimensions. For any practical usage of QPP, we recommend a set of 256 permutation matrices of a dimension more or equal to 2048. 

\end{abstract}

\maketitle
\section{Introduction}
In 1994, Peter Shor made a significant breakthrough by discovering quantum algorithms that operate in polynomial time for integer factorization and discrete logarithms \cite{Shor94}. The implication of this discovery is that if quantum computers of a sufficient scale are ever built, Shor's algorithms would entirely shatter RSA cryptosystem and signature scheme, Diffie-Hellman key exchange, and elliptic curve cryptography, including elliptic curve and elliptic curve digital signature algorithms \cite{Bernstein17}. It is worth noting that no such quantum computer currently exists to carry out cryptanalysis, but there has been significant progress made in recent years on both quantum computational hardware and algorithms.

The advancements in quantum computing have spurred a growing interest in quantum cryptography. This term can be used to refer to two distinct methods: Quantum Key Distribution (QKD) \cite{Pirandola20} and Post-Quantum Cryptography (PQC) \cite{Bernstein17}. Both of these techniques are utilized to create a secret key for digital symmetric encryption. QKD is based on principles of quantum mechanics and promises to offer information-theoretic security. PQC, on the other hand, utilizes classical algorithms with underlying mathematical problems that are conjectured to be unsolvable by quantum computing systems \cite{Wolf21, Seito21, Maimut19}. The primary objective of PQC is to develop cryptographic primitives that can resist quantum adversaries, i.e. adversaries with access to a quantum computer.

Kuang and Bettenburg introduced a new symmetric cryptographic algorithm in 2020 known as Quantum Permutation Pad (QPP) \cite{Kuang20}. The QPP algorithm uses permutation matrices to encrypt plaintext, and their corresponding Hermitian conjugates to decrypt the resulting ciphertext and reveal the original plaintext. To ensure that the decrypting party correctly applies the respective conjugates of the permutation matrices used for encryption, the two communicating parties pre-share a secret key using some asymmetric key exchange scheme. QPP has been used to develop lightweight block cipher \cite{Kuang21}, entropy expansion \cite{Lou21, avval22}, universal whitening algorithm \cite{gupta22} and pseudo-Quantum Random Number Generator or pQRNG \cite{RKuang21}. Kuang and Barbeau also proposed the concept of universal cryptography using QPP, which can be used in both classical and quantum computing systems \cite{Kuang22}. 

Although unconditional security proof exists for a number of QKD protocols, security against side-channel attacks is a work in progress \cite{Diamanti16, Sun16}. Similarly, the current cryptanalysis efforts have resulted in several loopholes in even the recommended algorithms by NIST for post-quantum cryptography \cite{Sounds21, Wouter22, Ward22, NIST}. The level of confidence we have in quantum cryptographic protocols is directly linked to the amount of cryptanalysis performed on the primitives. This enables researchers to determine the security margin, which measures how far the construction is from being broken. In the classical world, we have a large and constantly updated cryptanalysis toolbox that provides solid evaluations of primitive security against classical adversaries. However, this is no longer the case when considering quantum adversaries in the post-quantum world. To ensure post-quantum security for current ciphers and design new, secure ciphers for the post-quantum world, we need to develop a complete cryptanalysis toolbox for quantum adversaries similar to what has been accomplished for the classical world. This is a crucial step in correctly evaluating post-quantum security and protecting our digital infrastructure in a post-quantum era.

In this work, we address cryptanalysis in the context of a quantum permutation pad for a very specific but severe scenario. We captured the instances where there is a possibility of plaintext being equal to the ciphertext in the two possible scenarios, 1. Complete permutation group. 2. Incomplete permutation group was used to perform encryption. We have found that scenario-2 has an exponential advantage in avoiding such instances (plaintext = ciphertext). We have used an image to visually showcase our findings and establish that scenario-2 is quantum-safe. The structure of the paper is as follows: We recapitulate the notion of correctness and Shannon's Perfect Secrecy in the Preliminary Section - (\ref{Prelim}). The idea of a quantum permutation pad is revisited in Section - (\ref{QPP}). The cryptanalysis results in scenario-1 and 2 are shown in Section - (\ref{main}). Finally, we conclude the paper with a summary and future scope of the present work in Section - (\ref{Conclusion}). 

\section{Preliminaries}
\label{Prelim}
Let us recapitulate some cryptographic definitions.\\ \\
\textbf{Symmetric Key Encryption Schemes:} These are also known as private key encryption schemes. Any symmetric key encryption scheme $S$ is described by a 3-tuple of algorithms $(Gen, Enc, Dec)$. \\The algorithm $Gen$ is the key-generation algorithm that takes as input a security parameter and outputs a secret key $k$. The algorithm $Enc$ is the encryption algorithm that takes as inputs the secret key $k$ and a message (plaintext) $m$ and outputs a ciphertext $c$.
Lastly, the algorithm $Dec$ is a decryption algorithm that takes inputs from the ciphertext $c$ and the secret key $k$ and gives back the original plaintext $m$. Note that the same secret key is used for both encryption and decryption and hence the name `Symmetric Key Encryption'.\\ \\
\textbf{Correctness:} A symmetric key encryption scheme is said to be correct if the following condition holds-\\
\begin{equation*}
    Pr[m=Dec(k,c) | k\leftarrow Gen(\cdot), c\leftarrow Enc(k,m)] = 1
\end{equation*}
This means that given any ciphertext that was generated by a correct symmetric encryption scheme and the secret key that was used for encryption, the decryption algorithm of that scheme must always output the plaintext from which the ciphertext was generated. Quantum permutation pad (QPP) holds the correctness property.\\ \\
\textbf{Shannon Perfect Secrecy:} An encryption scheme provides `perfect secrecy' against a ciphertext-only attack (an attack in which the eavesdropper is only assumed to have access to the ciphertext and nothing else) if the attack provides no information about the plaintext even with full information about the ciphertext. In other words, if the plaintext and ciphertext are considered to be random variables, then they are independent or that:\\
\begin{center}
    \textit{Pr}[Guessing Plaintext | Ciphertext is Known] = \textit{Pr}[Guessing Plaintext]
\end{center}
QPP has been shown to hold Shannon Perfect Secrecy with the assumption that ciphertext is the same as plaintext in the trivial scenario only. However, we will show that there could be several such non-trivial instances in the case of QPP.  
\textbf{One Time Pad (OTP):} OTP is a symmetric key encryption scheme that offers Shannon Perfect Secrecy. The message space for OTP is the set of all binary strings. It generates a secret key of the same size as that of the plaintext. This secret key is a random binary string that is generated by a Binary Symmetric Source (BSS) (e.g. a perfectly fair coin in which heads represent `1' and tails represent `0' can be considered a BSS). The ciphertext is generated by taking a bitwise XOR of the plaintext and the secret key. The decryption algorithm works by again taking the bitwise XOR of the secret key and the ciphertext to give back the plaintext. OTP has been proven to be secure given the condition that the secret key is used only once.
\section{Quantum Permutation Pad}\label{QPP}
In this section, we will discuss a variant of the quantum permutation pad proposed by Kuang et al. in 2020 \cite{Kuang20}. Our variant allows us to do cryptanalysis on the behavior of Permutation matrices that are the fundamental building block of this cryptographic algorithm. This scheme is a symmetric key encryption scheme.\\

\subsection{Overview}
 The quantum permutation pad uses randomly selected permutation matrices to transform the input data in chunks of the size of the permutation matrix to produce the ciphertext. A permutation matrix is a square matrix in which each row and each column has exactly one non-zero element which is equal to 1. Multiplication of a vector by a permutation matrix shuffles the contents of that vector. Each chunk of the input is multiplied by a randomly selected permutation matrix (from a pre-generated set) and the permuted chunks are concatenated to give the output. The QPP scheme contains mainly three parts:
\subsection{Key Generation}
The secret 'key' here is composed of two parts - 1. Generation of the subset of the permutation matrices (transpose permutation matrices). 2. For a given instance, selection of the specific permutation matrix from the subset to be used for the encryption/ decryption. Note that we can never work with the complete set of permutation matrices as the size of the complete set is too big to be practically worked with. This, as we will see later, is actually beneficial for us. \\
The first part of the secret key is generated by repeated application of the Fisher-Yates shuffle algorithm (this part is done before the encryption algorithm starts working).  After this happens we can start the second part of the key generation as well as the encryption. Note that the permutation matrices generated here will be present in a certain order and thus can be assumed to be indexed (from $0$ to $m-1$ if $m$ matrices are chosen). This first part of the secret key gives the receiver information about the subset of permutation matrices using which the encryption was done. The receiver can use the same secret key to generate the transpose of the permutation matrix for the decryption procedure.
\subsection{Encryption }
For this part, we will consider chunks of the input data (of size equal to $N$ where the size of the permutation matrices is $N\times N$) sequentially as they are present in the input file. For each chunk, a permutation matrix will be chosen. The index of this matrix (again from $0$ to $m-1$ if $m$ matrices are chosen) can be randomly chosen. \\

Now, the index which is decided on goes into the second part of the secret key which will be used to give the receiver information about which matrix was used for which chunk. This process is repeated for every chunk of the input and at the end of this stage we have both the secret key as well as the ciphertext ready for transmission. The algorithm for the encryption mechanism is mentioned below:\\

$\bullet$ \textbf{Key Generation}
\begin{algorithmic}[1]
\State $i \leftarrow 1$
\While{$i \leq N+m$}
    \State $G[i] \leftarrow$ RandomInt(1,$N$)
    \State $M[i] \leftarrow$ RandomInt(0,$m-1$)
\EndWhile
\end{algorithmic}

$\bullet$\textbf{Permutation Matrix Generation Using Fisher-Yates Shuffle Algorithm:}\\
\begin{algorithmic}[1]
\Require RandomInt(1,$N$) returns a random integer between 1 and $N$ (inclusive) when called; Swap($a$,$b$) swaps the values of $a$ and $b$
\Ensure $P$ is a random permutation matrix
\State $i \leftarrow 1$
\While{$i \leq N$}
\State $K[i] \leftarrow$ $G[i]$
  \State $S[i] \leftarrow i$
  \State $j \leftarrow 1$
    \While{$j \leq N$}
        \State $P[i][j] \leftarrow 0$
        \State $j \leftarrow (j+1)$
    \EndWhile
    \State $i \leftarrow (i+1)$
\EndWhile
\State $i\leftarrow N$
\While{$i>0$}
    \State $p \leftarrow K[i]$
    \State $Swap(S[p],S[i])$
    \State $i \leftarrow (i-1)$
\EndWhile
\State $i \leftarrow 1$
\While{$i\leq N$}
    \State $P[i][S[i]] \leftarrow 1$
    \State $i \leftarrow (i+1)$
\EndWhile
\end{algorithmic}

$\bullet$ \textbf{Encryption}
\begin{algorithmic}[1]
\Require Dividing the input plaintext in chunks of $N$, say, $PT[i]_{i=1}^{i=N}$ (size of the permutation matrix)
\State $i \leftarrow 1$
\While{$i \leq \text{end of file}$}
    \State $M[i] \leftarrow M[i]$
    \State $j \leftarrow 1$
    \State $k \leftarrow 1$
    \While{$j \leq N$}
        \While{$k \leq N$}
           \State CT[j] += $P[j][k]*PT[k]$
            \State $k \leftarrow (k+1)$
        \EndWhile
        \State $j \leftarrow (j+1)$
    \EndWhile
    \State $i \leftarrow (i+1)$
\EndWhile    
\end{algorithmic}

\subsection{Decryption}
After the receiver has the secret keys (both the parts) and the ciphertext, they can proceed to decrypt the ciphertext using the decryption process. The main fact that will be useful for decryption is the orthogonality of permutation matrices - if $P$ is a permutation matrix and $P^T$ is its transpose then $PP^T = I$ where $I$ is the identity matrix.\\
The receiver has the information about which matrix was used to encode each chunk, hence they can break the ciphertext into chunks and multiply the transpose of those matrices with the output chunks to give back the input chunks. These decrypted chunks can be concatenated to give back the original plaintext.

$\bullet$ \textbf{Secret Key Retrieval at Receiver}
\begin{algorithmic}[1]
\Require Receiving entity gets the secret using some post-quantum cryptographic mechanism.
\end{algorithmic}

\textbf{Permutation Matrix Generation Using Fisher-Yates Shuffle Algorithm:}\\
\begin{algorithmic}[1]
\Require RandomInt(1,$N$) returns a random integer between 1 and $N$ (inclusive) when called; Swap($a$,$b$) swaps the values of $a$ and $b$
\Ensure $P$ is a random permutation matrix
\State $i \leftarrow 1$
\While{$i \leq N$}
\State $K[i] \leftarrow$ $G[i]$
  \State $S[i] \leftarrow i$
  \State $j \leftarrow 1$
    \While{$j \leq N$}
        \State $P[i][j] \leftarrow 0$
        \State $j \leftarrow (j+1)$
    \EndWhile
    \State $i \leftarrow (i+1)$
\EndWhile
\State $i\leftarrow N$
\While{$i>0$}
    \State $p \leftarrow K[i]$
    \State $Swap(S[p],S[i])$
    \State $i \leftarrow (i-1)$
\EndWhile
\State $i \leftarrow 1$
\While{$i\leq N$}
    \State $P[i][S[i]] \leftarrow 1$
    \State $i \leftarrow (i+1)$
\EndWhile
\Require Transpose of the generated permutation matrix ($Pt$).
\end{algorithmic}

$\bullet$ \textbf{Decryption}
\begin{algorithmic}[1]
\Require Dividing the received ciphertext in chunks of $N$, say, $CT[i]_{i=1}^{i=N}$ (size of the permutation matrix)
\State $i \leftarrow 1$
\While{$i \leq \text{end of file}$}
    \State $M[i] \leftarrow M[i]$
    \State $j \leftarrow 1$
    \State $k \leftarrow 1$
    \While{$j \leq N$}
        \While{$k \leq N$}
               \State PT[j] += $Pt[j][k]*CT[k]$
            \State $k \leftarrow (k+1)$
        \EndWhile
        \State $j \leftarrow (j+1)$
    \EndWhile
    \State $i \leftarrow (i+1)$
\EndWhile    
\end{algorithmic}

\section{Cryptanalysis}\label{main}
We analyzed the efficacy of the proposed quantum permutation pad and encounter a severe issue that can arise when using permutation matrices to encrypt a stream of bits that remain the same even after the permutation, the plaintext chunk may remain unchanged. This can happen if the permutation matrix chosen only permutes the bits which are 1 among themselves and the bits which are 0 among themselves. We now present a mathematical analysis to report how often these collisions do occur (a collision refers to a plaintext chunk that equals its ciphertext version even after multiplication with a permutation matrix).\\ 

\textbf{Scenario-1:} \textit{When a complete group of permutation matrices forms QPP}

In this case, we'll be assuming that we are working with the complete group of permutation matrices i.e. if we choose the size of the chunks to be $N$ bits, then we have access to all $N!$ permutation matrices (note that this is not a realistic assumption, as $N!$ gets very big very fast, this is only for analysis's sake - we will replace this with a more realistic assumption in the next section).\\

\textbf{Result-1:} \textit{In this scenario, the probability for a collision (ciphertext = plaintext) is inversely proportional to the dimension of the permutation matrix.}

\textbf{Proof:} Consider a single chunk of data ($N$ bits long). Let the number of bits in this chunk which are equal to $1$ be $p$, and the remaining $N-p$ bits be $0$. We will further assume that $1 \leq p \leq (N-1)$. This is because if $p=0$ or $p=N$, then the chunk is all zeroes or all ones respectively, and no permutation matrix will have an effect on this chunk. As these chunks are rare in real-world data, and they don't carry much meaningful information, we can safely assume this. \\
As we are working with the complete group of matrices, for the encryption, we first choose a matrix randomly from the available $N!$ matrices, and then multiply the plaintext chunk with this matrix. Note that there will be exactly $p! (N-p)!$ matrices out of the $N!$ which will not change the plaintext chunk after multiplication with them. This is because this is precisely the number of matrices that have the positions of the $p$ 1's in the chunk permuted among themselves, and the $(N-p)$ 0's also permuted among themselves. The probability of choosing such a matrix is thus -
\begin{equation*}
    P(plaintext = ciphertext) = \frac{p!(N-p)!}{N!} = \frac{1}{^NC_p}
\end{equation*}
But, as we know that $1 \leq p \leq (N-1)$, we can conclude that $^NC_p \geq N$. This implies that - 
\begin{equation*}
    P(plaintext = ciphertext) \leq \frac{1}{N}
\end{equation*}
If we choose chunks to be of sizes around 4096 (which is completely reasonable), we can see that the probability of collision even when working with the complete group of permutation matrices is less than $0.025\%$\\

\begin{figure*}
\centering
    \begin{subfigure}{8cm}
    \centering\includegraphics[width=8cm]{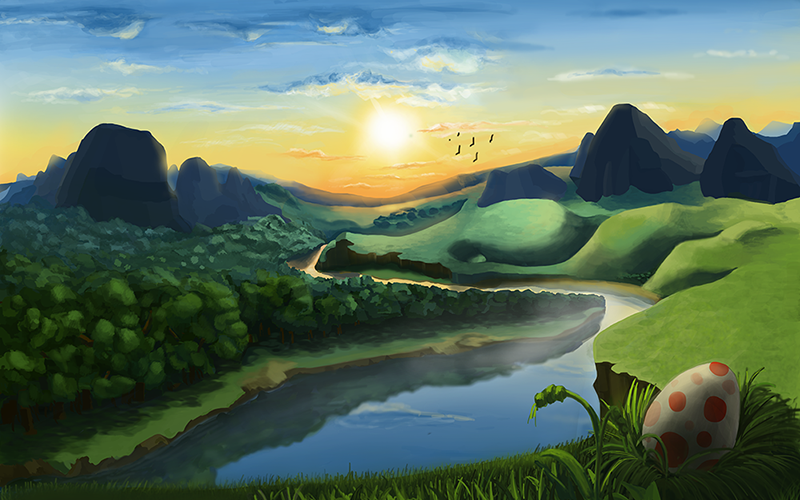}
    \caption{Original Image}
    \end{subfigure}%
    \begin{subfigure}{8cm}
    \centering\includegraphics[width=8cm]{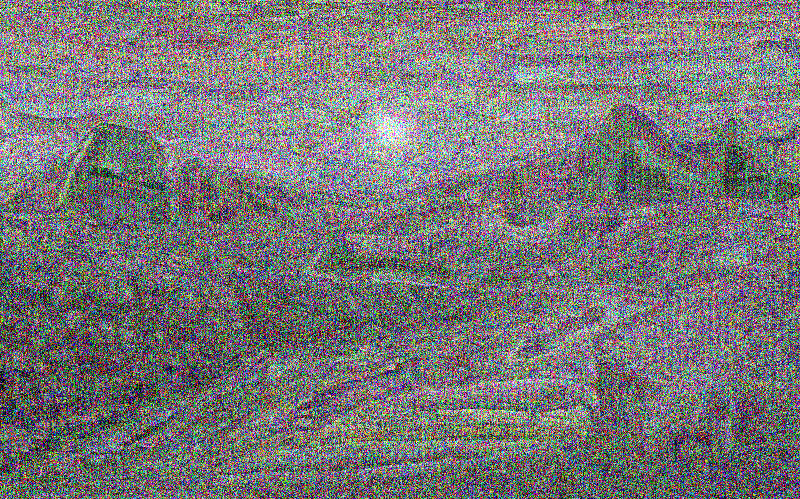}
    \caption{Cipherimage generated using $64\times 64$ Size Matrices}
    \end{subfigure}
    \begin{subfigure}{8cm}
    \centering\includegraphics[width=8cm]{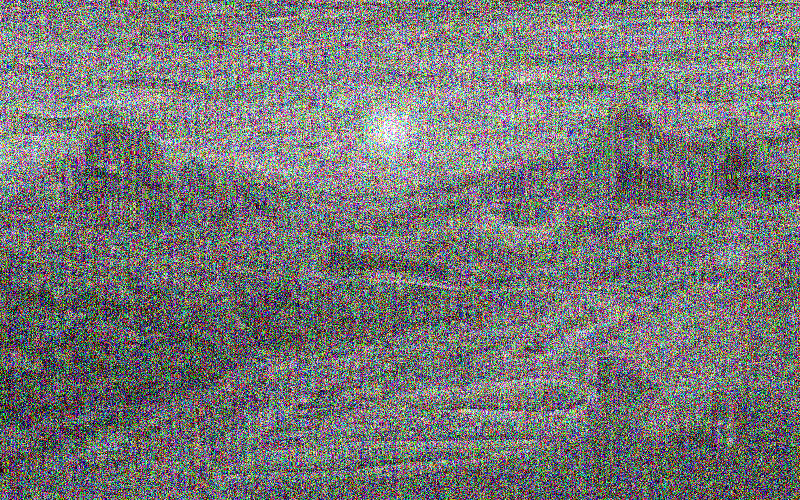}
    \caption{Cipherimage generated using $256\times 256$ Size Matrices}
    \end{subfigure}%
    \begin{subfigure}{8cm}
    \centering\includegraphics[width=8cm]{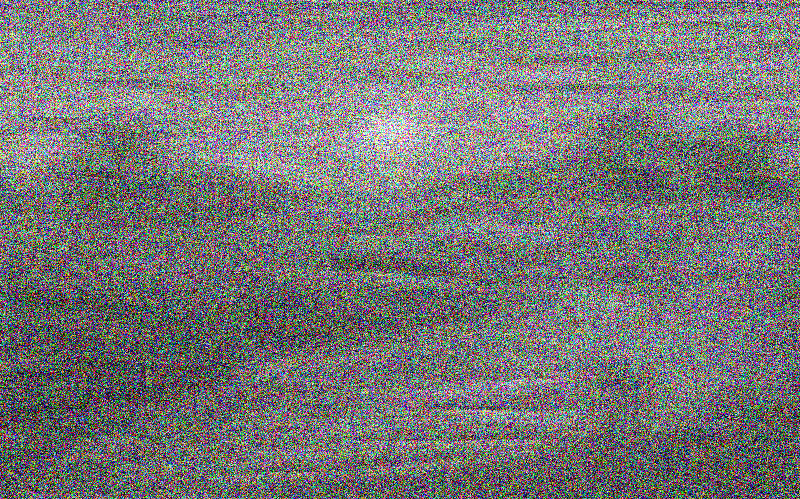}
    \caption{Cipherimage generated using $1024\times 1024$ Size Matrices}
    \end{subfigure}
    \begin{subfigure}{8cm}
    \centering\includegraphics[width=8cm]{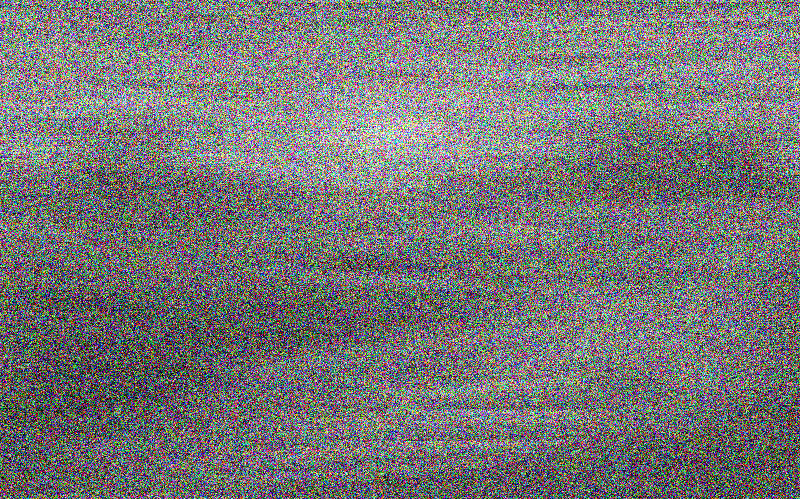}
    \caption{Cipherimage generated using $2048\times 2048$ Size Matrices}
    \end{subfigure}
    \begin{subfigure}{8cm}
    \centering\includegraphics[width=8cm]{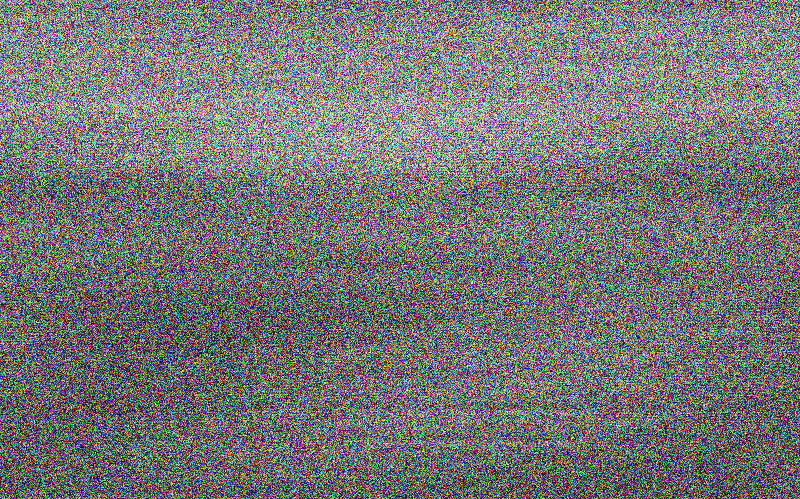}
    \caption{Cipherimage generated using $8192\times 8192$ Size Matrices}
    \end{subfigure}%
\caption{Collision effects (image impressions left) after the action of quantum permutation pad when 256 permutation matrices of different dimensions (64, 256, 1024, 2048, and 8192) are used.}
\label{scene}
\end{figure*}

\textbf{Scenario-2:} \textit{Collisions with Incomplete Group of Matrices}
Now, we relax the assumption of working with the complete group of matrices that we had in the previous section. Note that this is a more practical setting as we can only pre-generate a small number (as compared to the total number of permutation matrices) of permutation matrices. Let the number of matrices that we pre-generate prior to the starting of the encryption be $m$. For every chunk, we will choose a matrix out of these $m$ matrices at random. Another realistic assumption is that $m << (N-1)!$ (e.g. the case of 256 matrices of size 2048 falls in this case as 256 << (2048-1)! Similarly other practical cases also fall under this case). \\

\textbf{Result-2:} \textit{In this scenario, the probability of a collision (ciphertext = plaintext) is inversely proportional to the $m^{\text{th}}$ power of the dimension of the permutation matrix. Here, $m$ is the number of permutation matrices that form the QPP.}

\textbf{Proof:} Now, just like the previous chunk, we consider the encryption of a chunk of size $N$ with $p$ bits equal to 1 and the remaining $(N-p)$ equal to 0 (and the added assumption that $1 \leq p \leq (N-1))$. The worst case for encryption for this particular chunk will be the case in which all of the $m$ matrices on multiplication with the chunk don't change the chunk. We will consider this worst case probability. Again, note that out of the complete group of $N!$ matrices, there is $p!(N-p)!$ matrices that give the same ciphertext chunk as this particular plaintext chunk. \\
For all the $m$ matrices to not change the plaintext chunk when they are multiplied with it, all $m$ of them must be out of the $p!(N-p)!$ matrices. The probability of this happening when we randomly pre-generated the matrices is -
\begin{equation*}
    P(plaintext=ciphertext) = \frac{^{p!(N-p)!}C_m}{^{N!}C_m}
\end{equation*}
We can see that the numerator $^{p! (N-p)!}C_m$ is maximum when $p! (N-p)!$ is maximum. This will happen when $p = 1$ or $p = (N-1)$. Putting $p=1$ we get that -
\begin{equation*}
    P(plaintext=ciphertext)\leq \frac{^{(N-1)!}C_m}{^{N!}C_m}
\end{equation*}
On expanding the right-hand side, we get - 
\begin{equation*}
    \frac{^{(N-1)!}C_m}{^{N!}C_m} = \frac{\prod_{i=0}^{m-1}((N-1)!-i)}{\prod_{i=0}^{m-1}(N!-i)}
\end{equation*}
Now, by our assumption $m<<(N-1)!$, and so for all the terms in the numerator as well as the denominator, we can neglect $i$ wrt $(N-1)!$ and $N!$. This gives us-
\begin{equation*}
    \frac{^{(N-1)!}C_m}{^{N!}C_m} \approx \frac{\prod_{i=0}^{m-1}((N-1)!)}{\prod_{i=0}^{m-1}(N!)} = \frac{1}{N^m}
\end{equation*}
From the above, we can conclude that -
\begin{equation*}
    P(plaintext=ciphertext) \leq \frac{1}{N^m}
\end{equation*}
For practical use cases e.g. $N = 2048$, $m = 256$, the number $2048^{256}$ is so astronomically large that collision-free encryption is practically guaranteed as shown in Figure-{\ref{scene}}. Comparing this with the scenario when the entire permutation matrix group was considered, we see that the choice of pre-generating a fixed set of permutation matrices and using those is practical as well as more secure (in terms of lower collision probability) than working with the complete set of permutation matrices.

Using Figure-{\ref{scene}}, we have highlighted the severity of such collisions by showing the action of QPP on an image file. One can notice that the impression of the image is visible when the action of QPP was done using 256 matrices of dimensions 64, 256, and even 1024. However, as we move to even higher dimensions like 2048 and 8192, the image impressions are negligible. Note that the impressions are coming because some of the pixel information witness the instance of collision. Hence, it is advisable to always use higher dimensional permutation matrices (2048 or above) for the quantum permutation pad.  
 
\section{Conclusions}\label{Conclusion}
Cryptanalysis is an essential requirement to increase the confidence level that we have in quantum cryptographic protocols. We highlight the immediate requirement for a constantly updated cryptanalysis toolbox against quantum adversaries for quantum-safe cryptographic algorithms. This is a crucial step in correctly evaluating post-quantum security and protecting our digital infrastructure in a post-quantum era.

We have encountered instances of ciphertext being the same as plaintext even after the action of the proposed quantum permutation pad. We have examined such instances reported as collisions for two scenarios: 1. Complete set of permutation matrices is used for a given dimension. 2. Incomplete set of permutation matrices is used for a given dimension. We proved that the scenario of an incomplete set is advantageous to avoid such collisions where the probability of such instances is at most $\frac{1}{N^m}$. On the other hand, the collision probability is at most $\frac{1}{N}$ in the scenario of a complete set. We have also visually highlighted this issue by taking an image as plaintext and found the impression of the image left in the cipher image when the dimension of the permutation matrix used is 64, 256, and 1024. However, as we increase the dimension to 2048 or 8192, the impressions are negligible. Hence, a higher dimension permutation matrix (2048 or above) is recommended for any practical usage of the quantum permutation pad.

There are several offshoots of the present work. We have analyzed the security of QPP for the sharp instances where the plaintext is the same as the ciphertext. However, there are other instances that might be useful for eavesdropping. Determining other vulnerabilities will further strengthen the QPP and other quantum-safe algorithms in the longer run. 

\section*{Acknowledgement} 
SG acknowledges the financial support from QuNu Labs Pvt Ltd.

\appendix
\onecolumngrid


\begin{thebibliography}{99}

\bibitem{Shor94} Peter W. Shor, \emph{Polynomial-Time Algorithms for Prime Factorization and Discrete Logarithms on a Quantum Computer}, \href{https://doi.org/10.1137/s0097539795293172}{SIAM J.Sci.Statist.Comput. 26 (1997) 1484.}

\bibitem{Bernstein17} Bernstein, D., Lange, T. \emph{Post-quantum cryptography}, \href{https://www.nature.com/articles/nature23461}{Nature 549, 188–194 (2017)}

\bibitem{Pirandola20} S. Pirandola and U. L. Andersen and L. Banchi and M. Berta and D. Bunandar and R. Colbeck and D. Englund and T. Gehring and C. Lupo and C. Ottaviani and J. L. Pereira and M. Razavi and J. Shamsul Shaari and M. Tomamichel and V. C. Usenko and G. Vallone and P. Villoresi and P. Wallden, \emph{Advances in quantum cryptography}, \href{https://opg.optica.org/aop/abstract.cfm?uri=aop-12-4-1012}{Advances in Optics and Photonics Vol. 12, Issue 4, pp. 1012-1236 (2020)}

\bibitem{Gisin2002} Nicolas Gisin, Grégoire Ribordy, Wolfgang Tittel, and Hugo Zbinden, \emph{Quantum cryptography}, \href{https://link.aps.org/doi/10.1103/RevModPhys.74.145}{Rev. Mod. Phys. 74, 145 (2002).}

\bibitem{Wolf21}  Wolf R., \emph{Quantum key distribution: an introduction with exercises}, \href{https://link.springer.com/book/10.1007/978-3-030-73991-1}{Lecture notes in physics. vol. 988. Cham: Springer:2021}

\bibitem{Seito21} Seito T, Shikata J, \emph{Trend survey on post-quantum cryptography and its standardization}, \href{https://www.jstage.jst.go.jp/article/isciesci/65/2/65_60/_article/-char/ja/}{Syst Control Inf.
65(2):60–6 (2021)}

\bibitem{Maimut19} Maimut D, Simion E, \emph{Post-quantum cryptography and a (qu)bit more. In: Innovative security solutions for information technology and communications}, \href{https://link.springer.com/chapter/10.1007/978-3-030-12942-2_3}{Lecture notes in computer science. Cham: Springer; 2019. p. 22–8}


 




23. Kuang R, Barbeau M. Quantum permutation pad for universal quantum-safe cryptography. Quantum Inf Process.
2022;21:211

\bibitem{Kuang20} Randy Kuang and Nicolas Bettenburg, \emph{Shannon Perfect Secrecy in a Discrete Hilbert Space}, \href{https://ieeexplore.ieee.org/document/9259969}{IEEE International Conference on Quantum Computing and Engineering, QCE 2020, Denver, CO, USA, October 12-16, pp. 249-255, (2020)}

\bibitem{Kuang21} Kuang R, Lou D, He A, Conlon A., \emph{Quantum safe lightweight cryptography with quantum permutation pad}, \href{https://ieeexplore.ieee.org/document/9449247}{IEEE 6th International Conference on Computer and Communication Systems (ICCCS). 2021. p. 790–5 (2021)}


\bibitem{Lou21} Lou D, Kuang R, He A, \emph{Entropy transformation and expansion with quantum permutation pad for 5g secure networks}, \href{https://ieeexplore.ieee.org/document/9657891}{IEEE 21st International Conference on Communication Technology (ICCT) p. 840–5 (2021)}

\bibitem{gupta22} Avval Amil, Shashank Gupta, \emph{A universal whitening algorithm for commercial random number generators}, \href{https://arxiv.org/abs/2208.11935}{	arXiv:2208.11935 (2022)}

\bibitem{avval22} Avval Amil, Shashank Gupta,  \emph{Quantum entropy expansion using n-qubit permutation matrices in Galois field}, \href{https://arxiv.org/abs/2207.07148}{arXiv:2207.07148 (2022)}

\bibitem{RKuang21} Kuang R, Lou D, He A, McKenzie C, Redding M, \emph{Pseudo quantum random number generator with quantum
permutation pad}, \href{https://ieeexplore.ieee.org/document/9605294}{IEEE international conference on Quantum Computing and Engineering (QCE) (2021)}

\bibitem{Kuang22} Randy Kuang and Michel Barbeau, \emph{Quantum permutation pad for universal quantum-safe cryptography}, \href{https://link.springer.com/article/10.1007/s11128-022-03557-y#citeas}{Quantum Information Processing volume 21, Article number: 211 (2022)}

\bibitem{Diamanti16} Diamanti, E., Lo, HK., Qi, B. et al. \emph{Practical challenges in quantum key distribution}, \href{https://www.nature.com/articles/npjqi201625}{npj Quantum Inf 2, 16025 (2016)}

\bibitem{Sun16} Sun S, Huang A. \emph{A Review of Security Evaluation of Practical Quantum Key Distribution System}, \href{https://www.ncbi.nlm.nih.gov/pmc/articles/PMC8870823/}{Entropy (Basel). 2022 Feb 10;24(2):260 (2016)}

\bibitem{Sounds21} Soundes Marzougui, Nils Wisiol, Patrick Gersch, Juliane Krämer, Jean-Pierre Seifert, \emph{Machine-Learning Side-Channel Attacks on the GALACTICS Constant-Time Implementation of BLISS}, \href{https://arxiv.org/abs/2109.09461}{	arXiv:2109.09461 (2021)}

\bibitem{Wouter22} Wouter Castryck, Thomas Decru, \emph{An efficient key recovery attack on SIDH}, \href{https://eprint.iacr.org/2022/975}{Cryptology ePrint Archive, Paper 2022/975 (2022)}

\bibitem{Ward22} Ward Beullens, \emph{Breaking Rainbow Takes a Weekend on a Laptop}, \href{https://eprint.iacr.org/2022/214}{Cryptology ePrint Archive, Paper 2022/214 (2022)}

\bibitem{NIST} National  Institute  of  Standards  and  Technology, \emph{NIST  computer security resource center (CSRC)}, \href{http://csrc.nist.gov/groups/ST/toolkit/rng/index.html}{NIST, (2022)}


\end{thebibliography}
\end{document}